# A Novel Image-centric Approach Towards Direct Volume Rendering

NAIMUL MEFRAZ KHAN, Ryerson University
RIADH KSANTINI, University of Windsor, SUP'COM
LING GUAN, Ryerson University

Transfer Function (TF) generation is a fundamental problem in Direct Volume Rendering (DVR). A TF maps voxels to color and opacity values to reveal inner structures. Existing TF tools are complex and unintuitive for the users who are more likely to be medical professionals than computer scientists. In this paper, we propose a novel image-centric method for TF generation where instead of complex tools, the user directly manipulates volume data to generate DVR. The user's work is further simplified by presenting only the most informative volume slices for selection. Based on the selected parts, the voxels are classified using our novel Sparse Nonparametric Support Vector Machine classifier, which combines both local and near-global distributional information of the training data. The voxel classes are mapped to aesthetically pleasing and distinguishable color and opacity values using harmonic colors. Experimental results on several benchmark datasets and a detailed user survey show the effectiveness of the proposed method.

## 1. INTRODUCTION

Direct Volume Rendering (DVR) is a technique to reveal interesting structures and regions from raw 3D imaging data, typically obtained through popular medical imaging procedures such as Magnetic Resonance Imaging (MRI) and Computed Tomography (CT) [Kindlmann and Durkin 1998]. DVR makes use of a Transfer Function (TF), which maps one or more features extracted from the data (the feature space) to different optical properties such as color and opacity. The TF design is typically a user-controlled process, where the user interacts with different widgets (usually representing feature clusters or 1D/2D histograms) to manually set color and opacity properties to the feature space. In case of clustering-based TF design, the user can control some low-level properties like number of clusters, cluster variance etc. Most of the recently proposed DVR methodologies [Kindlmann and Durkin 1998; Nguyen et al. 2012; Wang et al. 2011; Selver et al. 2009] are based on these basics.

However, interacting with the feature space is difficult for the end-user, who may not have any knowledge about image processing or clustering. Multi-dimensional feature spaces cannot represent distinguishable properties such as peaks and valleys which are important for proper TF generation from histogram [Maciejewski et al. 2009]. Also, these kind of widgets try to represent the feature space directly, putting a strict restriction on the type of features used and the dimensionality.

In this paper, we propose a rather direct approach to simplify the process of volume visualization. Instead of working with complex widgets for histogram or cluster manipulation, the user simply works on the volume data itself. The user is presented with grayscale form of some slices from the volume data, where he/she can do simple selection on voxels to express his/her intention of how the volume should be classified. To further simplify the process, we carefully pick the most representative slices from the volume and only show those to the user. The slices are picked by sorting them based on image entropy, which provides a measure of information present in one slice [Studholme et al. 1999]. Once the user selection is completed, we treat the selected voxels as training data and extract some high-dimensional features. A recently proposed Sparse Nonparametric Support Vector Machine (SN-SVM) [Khan et al. 2014a] is then used to classify the whole volume. This approach combines the local information available through support vectors [Cristianini and Shawe-Taylor 2000] and the near-global information available through Kernel Nonparametric Discriminant (KND) [Mika et al. 1999] to provide an accurate high-dimensional classification. Due to the robustness of the classifier, only a small number of training samples can provide excellent results, as we will see from the experiments. After the voxel classification, the classes are mapped to different color and opacity values automatically by using the concept of color harmonization [Liu et al. 2011], which can generate easily distinguishable and aesthetically pleasing visualization of the underlying classes.

The rest of the paper is organized as follows: In the next section we present the related work. Section 3 describes the novel image-centric approach. Section 4 provides a comparative evaluation of our proposed image-centric approach to the data-centric approach proposed by [Khan et al. 2014b; Khan et al. 2015] and an off-the-shelf traditional TF editor [Kitware$^{TM}$ 2015]. Finally, Section 5 provides some conclusions.

## 2. RELATED WORK

Traditionally, TFs were one-dimensional, where the color and opacity are derived from intensity value only. The user assigns different color and opacity values to the one-dimensional histogram derived from intensity values [Levoy 1988]. However, volume data is typically complex in nature and naturally, intensity value alone is not a good feature to separate voxels into different groups.

Efforts have been taken to improve the TF generation process. The existing methods to generate TF for DVR can be divided into two categories: data-centric methods and image-centric methods.

— **Data-centric Methods**: In the *data-centric* approach, the system tries to find the TF solely based on data properties. Besides intensity, other properties of the data (e.g. gradient magnitude) can be used to design multi-dimensional TFs. The work of Kindlmann and Durkin first popularized multi-dimensional TF [Kindlmann and Durkin 1998]. Their work used gradient magnitude alongside intensity value to build a 2D histogram, where material boundaries in a volume can be identified as arches. The arches can be assigned color and opacity values manually with the help of the histogram. The limitation of this method was the overlapping of arches for different boundaries, making them indistinguishable. Their later work [Kniss et al. 2001a] has included the second derivative to build the histograms. In [Sereda et al. 2006a] the *LH-Histogram* was proposed. The LH-histogram is calculated by tracing the voxels along the boundary. The low and high intensity values along the gradient were recorded and used to build a histogram. In this approach, boundaries are represented as blobs rather than arches, which are relatively easier to distinguish [Sereda et al. 2006a]. The main drawback with this approach is that the histogram takes a long time to calculate [Sereda et al. 2006a].

Rather than using the magnitude of the voxel properties, the statistical relationships among them (mean, standard deviation etc.) were used to generate the histograms in [Maciejewski et al. 2013]. However, directly using statistical properties is sensitive to noise as mentioned by the authors in [Maciejewski et al. 2013].

Recently, *feature size* has also been used to design multi-dimensional TFs. In [Correa and Ma 2008], the voxels were assigned a scale field, which is calculated from the relative size of a particular feature for each voxel. A size-based distribution is then presented to the user for TF generation. A region-growing technique was used to find the feature size in [Hadwiger et al. 2008]. In [Wesarg et al. 2010], structure sizes were estimated based on a user-defined intensity range. This structure size was then used to generate a Structure Size Enhanced (SSE) histogram, which in turns was used to specify the TF.

Users generally may not be able to focus on voxels too far apart due to the complexity of volume datasets. As a result, spatial information has been used as features in some recently proposed methods. Local histograms combining intensity and spatial information in the local neighborhood was proposed in [Lundstrom et al. 2006a]. In [Lundstrom et al. 2006b], the $\alpha$-histogram was proposed, which incorporates the property of spatial coherence to produce a global histogram. Spatial information was combined with the 2D histogram in [Roettger et al. 2005]. A distance-based TF was introduced in [Tappenbeck et al. 2006] which assigns optical properties to structures based on the distance to a selected reference structure in order to hide or emphasize structures. A Volume Histogram Stack (VHS) is introduced in [Selver et al. 2009], where the histogram is created by aligning the histograms of the different slices and incorporating spatial information with them. Then a Self-Generating Hierarchical Radial Basis Function Network is used to generate the TF.

Generally speaking, the problem with histogram-based TF generation is that the method is restricted to use low-dimensional (typically upto 2D) features, since the feature space is represented directly with the histogram. Moreover, due to the inter-disciplinary use of volume rendering, end-users may not have deep understanding of image processing and computer graphics techniques i.e. histograms. Therefore, the complexity of a histogram-based interface can hinder their work flow. For these reasons, some recent methods have used different clustering and classification algorithms on the histogram data to divide the voxels into different groups and assign TF properties accordingly. Voxels were grouped based on their LH-histogram and spatial information in [Sereda et al. 2006b]. A non-parametric TF automation method based on kernel density estimation was proposed in [Maciejewski et al. 2009]. The estimated density distribution is divided into different ranges, and the user specifies the color and opacity values for each range to generate the final output. A parallel technique based on mean-shift clustering of voxel intensities and spatial values was proposed in [Zhou et al. 2010]. A semi-automatic method based on a combination of mean-shift clustering and hierarchical clustering was proposed in [Nguyen et al. 2012], where the user has control over the hierarchical clustering results. Self-Organizing Maps were used in [Pinto and Freitas 2007] to perform dimensionality reduction on the feature space. Different features of the voxels and the maps were then linked to the color and opacity values to generate the final rendering based on user action. More recently, [Khan et al. 2014b; Khan et al. 2015] proposed a novel data-centric approach, where they generated an organized representation of the data through clustering and provide the user with some intuitive control over the output in the cluster domain. They used Spherical Self-Organizing Maps (SSOM) as the core of their approach. Instead of manipulating complex widgets, the user interacts with the simple SSOM color-coded lattice to design the TF. Even with these methods, the end user will still have to manipulate low level cluster parameters and assign the color and opacity values manually.

— **Image-centric Methods**: Another alternative for TF specification is the *image-centric* approach, where the users are given the option to interact with the data in the image domain i.e. directly with the image slices or the rendered volume itself [S. Fang et al. 1998]. The changes to the TF are done in the background. Image-centric approaches also include methods where the TF is picked by the user from a set of TFs based on the output rendering, not based on volume properties. The Design Gallery method [Marks et al. 1997] is a popular image-centric approach. In this method, several different TFs are generated by varying the input parameters. The user is then presented with the output generated from all the TFs, and the task of final TF selection is left to the user. In [Guo et al. 2014], a similar approach is used to explore TFs designed by many previous users. The user has the ability to look at the history of the previous TFs and interpolate from one to another to discover new features in the volume data and explore the different approaches previous users have taken to design TFs and the relation between them. Gaussian Mixture Models (GMM) were used in [Guo et al. 2011b], where the user can manipulate the different parameters of the models instead of directly manipulating TF parameters. An intelligent user interface has been introduced in [Tzeng 2006], where the user can paint on different slices to mark areas of interests. Neural network based techniques are then used to generate the TF. A spreadsheet-like interface based on Douglas-Peucker algorithm is presented in [Liu et al. 2010].

  The methods proposed in [Guo et al. 2011a; Guo and Yuan 2013] are also image-centric, where the user interacts directly with the rendered volume in a What-You-See-Is-What-You-Get (WYSIWYG) style. However, our system approaches the problem from a different angle. The most fundamental difference is that the method in [Guo and Yuan 2013] relies on major pre-processing of the raw volume data before any user interaction takes place. In the pre-processing phase, the method applies branch decomposition based on a contour tree computed from the raw volume data to segment it into a labeled volume. The static number of branches controls the initial semantic realization of the volume data and, consequently, the initial TF. As a result, the user may already be losing some information he/she was looking for. The initial parameters will heavily influence the exploration process. On the other hand, in our system, the user interacts with the raw slices directly. Medical professionals are quite familiar with the 2D interpretation of raw image slices despite their noisy nature, and interacting with the raw data provides them with a more granular level of control compared to the method proposed in [Guo and Yuan 2013]. Moreover, our method is possibly the first of its kind with an elaborate quantitative user survey that compares the proposed system with others. Most DVR method evaluations are qualitative, where the user simply expresses his/her opinion with words [Guo and Yuan 2013]. We have gone one step further and provided a detailed breakdown of how each component of the system affects the user's exploration process.

— **Color Assignment**: Most of the methods discussed above mainly focused on classifying the voxels in the volume data into different groups/regions. However, the final rendering is obtained by assigning some color and opacity values to the voxels. In most of the existing methods, this color and opacity assignment is manual. The visual aesthetics is very important for volume rendering, as the user can only draw required conclusions from the volume data visually. So the colors assigned to different voxel groups need to be differentiable. The concept of *color harmonization* can help here [Itten 1969]. Harmonic colors are a set of colors that look aesthetically pleasing when visualized together. Using harmonic colors can take the burden of color selection away from the user.

Zhou et al. were the first to use the concept of color harmonization in TF generation [Zhou and Takatsuka 2009]. Their method adopted a residue flow model based on Darcy's law for TF generation. Color harmonization was also used in [Liu et al. 2011], where the K-means++ algorithm was used to divide the voxels into groups based on their intensity values and spatial information. This method also uses automatic assignment of opacity values based on the user's perception.

Our proposed method approaches the problem from an *image-centric* viewpoint [Tzeng 2006; Kniss et al. 2001b], where the user directly performs operations on the most informative volume slices rather than on histograms or cluster visualizations. We also automate the color and opacity assignment based on color harmonization and user's perception and viewing habits. As a result, the user can focus on the interpretation of the result rather than manipulating complex interfaces.

## 3. PROPOSED METHOD

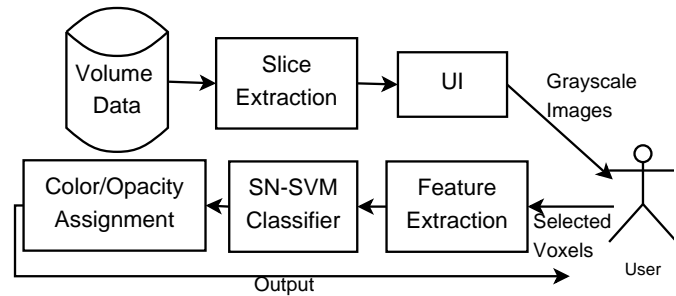

Fig. 1. System diagram for the proposed method.

Figure 1 shows the high level system diagram for the proposed method. As can be seen, in our image-centric approach, the user is first presented with grayscale form of some slices from the volume data, where he/she can do simple selection of voxels to express his/her intention of how the volume should be classified. To further simplify the process, we carefully pick the most representative slices from the volume and only show those to the user. The slices are picked by sorting them based on image entropy, which provides a measure of information present in one slice [Studholme et al. 1999]. Once the user selection is completed, we treat the selected voxels as training data and extract some high-dimensional features. Our proposed SN-SVM classifier is then used to classify the rest of the voxels into different groups. The combination of global and local distributional information results in an accurate classification. Also, due to the robustness of the classifier, only a small number of training samples can provide excellent results, as we will see from the experiments. After voxel classification, the classes are mapped to different color and opacity values automatically by using the concept of color harmonization [Liu et al. 2011]. This automated color generation scheme can generate easily distinguishable and aesthetically pleasing visualization of the underlying classes. The various modules of the proposed method are described in the next sections.

### 3.1. Slice Extraction
Since the user will perform selection operations on volume slices, we need to provide the user with the *most informative* slices. A typical volume can have thousands of slices (along X, Y or Z direction). We need to provide the user with the slices that contain the most variation, since we can safely assume that these slices will contain all the structures that the user might be interested in [Studholme et al. 1999]. For this, we calculate the *image entropy* of each slice. In general, for a set of $M$ symbols with probabilities $p_1, p_2, \ldots, p_M$ the entropy can be calculated as follows [Studholme et al. 1999]:

$$H = -\sum_{i=1}^{M} p_i \log p_i. \quad (1)$$

For an image (a single slice), the entropy can be similarly calculated from the histogram [Studholme et al. 1999]. The entropy provides a measure of variation in a slice. Hence, if we sort the slices in terms of entropy in descending order, the slices with the highest entropy values can be considered as representative slices.

### 3.2. User Input

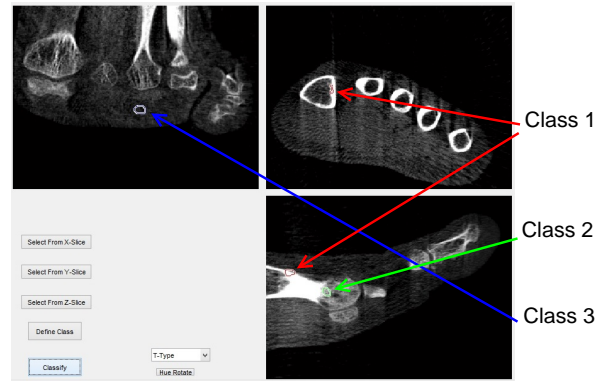

Fig. 2. The GUI with extracted slices. The red, green and blue selections (circles) by the user refer to voxels assigned to Class 1, Class 2, and Class 3, respectively.

The user is presented with a GUI (Figure 2), which shows the three slices with highest entropy values in X, Y and Z direction, respectively. The user can then select different voxels and assign them to different classes, which will be treated as training data for the next steps in the method. To select voxels to assign to a class, the user draws a free-form boundary around a group of voxels by clicking and dragging the mouse (or using his/her finger on a touchscreen). Once a boundary is drawn, all the voxels inside the drawn region are assigned to a voxel class by clicking "define class". As we can see from Figure 2, the regions assigned to different classes by the user are marked with different colors. The dataset shown here is the Foot CT (details can be found in Section 4). The objective is to roughly separate the bones (Class 1), joints (Class 2), and the outer layer (Class 3) of the Foot. Please note that although the number of training voxels seems small, our proposed SN-SVM classification method can classify the whole volume from a small training set reliably by combining both local and near-global distributional information [Khan et al. 2014a].

### 3.3. Feature Extraction

Unlike histogram-based methods, our approach does not restrict the feature dimension. Therefore, we can use a reasonably high-dimensional feature set. The features need to be picked carefully so that in the classifier stage, there is enough distinction between different classes. In the data-centric approach proposed by [Khan et al. 2014b;

Khan et al. 2015], the features being used were intensity, 3D gradient magnitude, the $X$, $Y$ and $Z$ locations of the voxels and the second derivative. However, we have found that using the same features for our image-centric approach did not provide satisfactory results. The reason behind this discrepancy is the way the voxels are being classified. In the data-centric approach [Khan et al. 2014b; Khan et al. 2015], the clustering process is unsupervised. But in the image-centric approach, the training voxels are picked by the user from the grayscale representation of the volume slices. Since the grayscale images are generated from intensity values alone, we have found that we need more emphasis on the local variances in intensity to capture the distinctive properties of each class. In other words, the intensity values of the neighboring voxels need to be taken into consideration. As a result, the average neighboring voxel values are used as separate features in our image-centric approach. We have also found that due to the strong localization capabilities of these intensity driven features, the $X$, $Y$, $Z$ locations and the second derivative value are not required to be included in the feature set to achieve satisfactory results. Dropping these redundant features, the 5D feature space for the proposed image-centric method consist of:

— the intensity value,
— the 3D gradient magnitude of each voxel, defined by:

$$G = \sqrt{G_x^2 + G_y^2 + G_z^2}, \tag{2}$$

where $G_x$, $G_y$ and $G_z$ are the gradient values along $X$, $Y$ and $Z$ direction, respectively.
— The average intensity value of the neighboring voxels. The intensity values of the neighboring voxels are divided into three groups by changing two of the three indices ($X$, $Y$ and $Z$) at a time while keeping the other one fixed. For example, if we keep the $z$ index fixed, the first group will consist of voxels with indices: $(x-1, y-1, z), (x-1, y, z), (x-1, y+1, z), (x, y-1, z), (x, y+1, z), (x+1, y-1, z), (x+1, y, z), (x+1, y+1, z)$. The three average values of these three different groups are treated as different features.

These features provide us with reasonable localization of voxel attributes, which helps separating different structures in the volume in the classification stage.

### 3.4. SN-SVM Classifier

The SN-SVM classifier [Khan et al. 2014a] is motivated by combining the merits of both discriminant-based classifier such as KND [Mika et al. 1999] and the classical SVM. The KND calculates the within-class scatter matrix by considering the $\kappa$-nearest neighbors for each training data point. Thus it considers the "near-global" characteristics of the training distribution. On the other hand, SVM only considers the "local" characteristics (support vectors) to build the separating hyperplane. Both of these sources of information are important for accuracy [Khan et al. 2014a]. In SN-SVM, these two are combined by incorporating the within-class scatter matrix $\Delta$ and the between-class scatter matrix $\nabla$ of KND into the SVM optimization problem. Let $\mathcal{X} = \{x_i\}_{i=1}^N$ represents the training data and $\mathcal{T} = \{t_i\}_{i=1}^N$ represents the associated class tags for a two-class problem. We also use the *kernel trick* [Cristianini and Shawe-Taylor 2000] to map the data points to a higher-dimensional feature space with the function $\Phi$. Then, the SN-SVM formulation can be described by the following optimization problem:

$$\min_{\mathbf{w}\neq 0, w_0} \left\{ \frac{1}{2}\mathbf{w}^T(\eta\Delta(\nabla+\beta I)^{-1}\Delta+I)\mathbf{w} \right.$$
$$\left. +C\sum_{i=1}^{N} max(0, 1-t_i(\Phi^T(x_i)\mathbf{w}+w_0)) \right\}. \quad (3)$$

Here, $\mathbf{w}$ and $w_0$ are the weight vector and the offset to be optimized. $\eta$ is the control parameter which dictates the amount of contribution from SVM and KND. By using an appropriate value of $\eta$, we can control the direction of the separating hyperplane of the classifier and place it in an optimum way. In [Khan et al. 2014a], we have also shown that the solution provided by SN-SVM is more sparse than the classical SVM, which can be utilized by efficient numerical methods to significantly speed up computation[Camps-Valls and Bruzzone 2005].

The value of $\eta$ is set to $0.3$ through experiments in the proposed system. Since the problem can be multi-class, we have used the one-vs-one classification scheme with our SN-SVM method [Khan et al. 2014a].

### 3.5. Color and Opacity Assignment

Once the classification is complete and the separate voxel classes are defined using the SN-SVM, we generate the color and opacity values to create the final rendering. We've put special attention to the color selection of different groups. In most of the existing methods, the colors are often determined by adhoc assignment or personal preference [Liu et al. 2011]. Since volume exploration is a lengthy and complex process, visual aesthetics is very important. The contrast among the colors is also important to differ among different types of materials (voxel groups in our case) efficiently. Hence, we apply the theory of color harmonization [Tokumaru et al. 2002] for our color assignment. Harmonic colors define sets of colors that are aesthetically pleasing to the user [Tokumaru et al. 2002]. In [Itten 1969], a color wheel was introduced in which color harmony is described across color hue values. A harmonic set is described by specifying the relative position of the colors on the wheel rather than specific color themselves. The harmonic colors are defined based on user evaluations rather than strict mathematical formulations. Based on this color wheel, several templates can be defined. Figure 3 shows the four templates we've used in our method, namely, Type V, Type L, Type T and Type X. All these templates can be rotated at arbitrary degrees to generate a new set of colors (please refer to [Cohen-Or et al. 2006] for the detailed specifications on these templates). We use these templates to generate the hue values for $HSV$ color space. We need separate hue value for each group of voxels selected by the user. We equally space the required hue values along the available hue range so that we have a visually pleasing result as a combination of all of them.

However, if we simply use one hue value for each user selected group, the details will not be revealed. For volume rendering, boundary enhancement is important to reveal the inner structures [Kindlmann and Durkin 1998]. As a result, instead of using one hue value, we use a small range with one lower and one higher limit, $\xi_l^i$ and $\xi_h^i$ for each group. For our experiments, the difference between $\xi_l^i$ and $\xi_h^i$ is $5\%$ of the whole hue wheel (or $18°$). This still gives us enough hue range to space out all the groups reasonably around the hue templates.

The hue value for each voxel of each group is calculated by the following equation:

$$H_i^v = \xi_l^i + (\xi_h^i - \xi_l^i) * F_v, \quad i = 1, \ldots, Z, \quad (4)$$

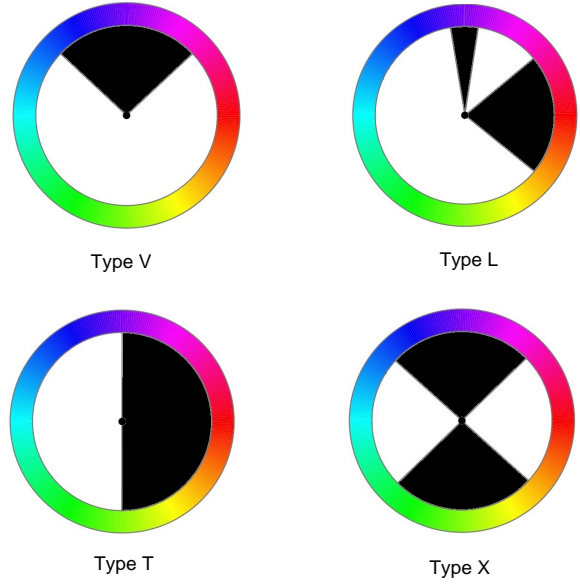

Fig. 3. Hue Templates

where $Z$ denotes the total number of groups selected by the user. $F_v$ is the *Gradient Factor* and is defined as:

$$F_v = \frac{G_v}{max_{v \in \mathcal{Z}} G_v}. \quad (5)$$

Here, $\mathcal{Z}$ denotes the group that the voxel $v$ belongs to. $G_v$ is the 3D gradient magnitude for each voxel, defined by:

$$G_v = \sqrt{G_x^2 + G_y^2 + G_z^2}, \quad (6)$$

where $G_x$, $G_y$ and $G_z$ are the gradient values along $X$, $Y$ and $Z$ direction, respectively.

In this way, even for a single group, the hue values will have some variation based on the gradient value instead of having a single hue value assigned to all of them.

In the $HSV$ color space, the $S$ and $V$ values are also very important to define a color. The $S$ and $V$ values respectively determine the saturation and brightness of the color. We generate the $S$ and $V$ values based on the understanding of the user's perception [Liu et al. 2011]. Voxel groups that have a small spatial variance occupy a smaller viewing area compared to groups that have relatively larger spatial variance. We can intuitively assume that the groups with smaller spatial variance needs to have a more saturated color to highlight them properly [Liu et al. 2011]. Hence, we calculate the $S$ value for each voxel group according to the following equation:

$$S_i = \frac{1}{(1 + \sigma_i^2)}, \quad i = 1, \ldots, Z. \quad (7)$$

Here, $\sigma_i^2$ represents the spatial variance of each voxel group.

Similarly, since the rendering results are usually viewed at a distance from the whole volume, we can assume that voxel groups closer to the center of the volume needs to be brighter so that they are not overshadowed by groups that are further from the center [Liu et al. 2011]. Hence, The $V$ value for each voxel group is calculated as follows:

$$V_i = \frac{1}{(1 + \mathbf{D}_i)}, \quad i = 1, \ldots, Z. \tag{8}$$

Here, $\mathbf{D}_i$ denotes the distance of the centroid of the $i-$th group to the center of the volume [Liu et al. 2011]. By assigning the saturation and brightness values this way, we can assign more vivid colors to the voxel groups that are relatively more difficult to highlight.

The last parameter to define the full TF is opacity. Since voxel groups with smaller spatial variances are likely to be obstructed for proper viewing by groups with larger spatial variances, we calculate the opacity values based on spatial variances. We also want to emphasize the boundaries to reveal the detailed structures. Hence, our opacity values for each voxel are calculated using the following equation:

$$O_i^v = \left(1 - \frac{\sigma_i^2}{\max(\sigma_i^2)}\right) * (1 + \mathcal{F}_v), \quad i = 1, \ldots, Z. \tag{9}$$

Here, $\sigma_i^2$ is the spatial variance of the $i-$th voxel group. $\mathcal{F}_v$ is the Gradient Factor as defined in Equation 5. Inspecting Equation 9, we can see that voxel groups with smaller spatial variances will be assigned higher opacity values. Moreover, all the opacity values will go through boundary enhancement by being multiplied by $(1+\mathcal{F}_v)$. The opacity values are scaled to fall between $0.2$ and $1.0$ to ensure that none of the voxel groups become completely transparent.

After calculating the $HSV$ and opacity values, we convert them into the final $RGBA$ texture. The $HSV$ triplets are converted into $RGB$, and the opacity value is appended to generate the $RGBA$ quadruples. This is passed to the rendering stage to generate the final rendering.

For the rendering stage, we use the Visualization Toolkit (VTK) [Schroeder et al. 2006]. The $RGBA$ texture passed to VTK is rendered in GPU. As a result, the user can see the rendering changes in real time while selecting and assigning voxels into different groups.

The color and opacity values are optimized with the assumption that the volume will be viewed as a whole. Typically, while examining a volume the user can perform cuts on the bounding box to separate out inner structure, or zoom into the render to examine specific voxels closely. The user may also want to emphasize certain voxel groups regardless of their relation to other voxel groups. In case of such changes in the viewing parameters, the color and opacity values will have to be updated dynamically to reflect the current set of voxels in view.

Also, a limitation of the rendering stage in this system is that it uses pre-classification DVR to speed up computational time by passing the RGBA texture directly to the GPU. State-of-the-art DVR pipeline uses post-classification for better visual results [Hadwigers et al. 2006], where the scalar voxel values are interpolated before classification. Adapting our proposed system to an existing post-classification DVR pipeline will require additional 3D texture to store group IDs for each voxel and maximum gradient magnitude for each group. This will result in slower rendering time due to the required additional dependent table lookups.

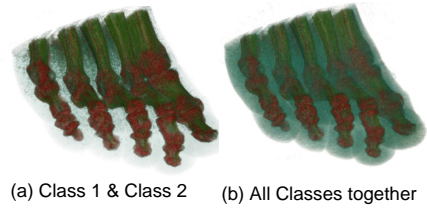

(a) Class 1 & Class 2    (b) All Classes together

Fig. 4. Results for the Foot dataset.

## 4. EXPERIMENTAL RESULTS

In this section, we provide an evaluation of our proposed image-centric approach, and we compare it to the data-centric approach proposed by [Khan et al. 2014b; Khan et al. 2015] and an off-the-shelf traditional TF editor [Kitware[TM] 2015] [1].

The off-the-shelf editor is VolView 3 by KitWare [Kitware[TM] 2015]. VolView only offers a 1D TF editor. Since our proposed system incorporates gradient values as features, comparison with a 2D TF editor would have been more ideal. However, we were unable to find an off-the-shelf 2D TF editor software that was user-friendly enough to compare against our system. Since it was critical that we provide the users with similar user-friendly response across all the systems being compared, we selected VolView. Although the TFE offered by VolView is 1D, editing the TF is very intuitive since it is professionally developed. Users can add points onto the 1D intensity histogram by simply clicking on it, and move the control points around through click-and-drag for easy changing of color and opacity mapping. Also, in general, 2D TFs are even more difficult to build by hand when compared with 1D TFs [Pfister et al. 2001].

VolView was configured to use pre-classification rendering for a faithful comparison with the proposed system.

### 4.1. Evaluation of the Proposed Image-Centric Approach

The same 5 datasets used in the data-centric approach [Khan et al. 2014b; Khan et al. 2015], are used to evaluate the proposed method. The datasets are: CT scans of a Foot, the Visual Male Head dataset, the Carp dataset, the Engine dataset and the Lobster dataset. Details of the datasets are shown in Table I for convenience. To speed up the classification process, we threshold these datasets with a value close to zero so that the air surrounding the actual data are not passed on to the classifier.

Figure 4 shows the result obtained based on the training data shown in Figure 2. Here, we see that our SN-SVM method can separate the bones, joints and the outer layer of the foot effectively. Figure 4-(b) shows that due to the use of intelligent color and opacity assignment, all three classes can be visualized at the same time and easily distinguishable.

Figure 5 shows comparison between SN-SVM and the classical SVM (based on the same training data). The bones and joints are shown together. We can see the advantage of SN-SVM when compared to SVM. In the areas pointed by arrows, the SVM was unable to accurately separate the joints from the bones, while the SN-SVM method was successful. This shows the superiority of the combined approach in SN-SVM. Although the SN-SVM result may look noisy in some areas, the target here is to show the effectiveness in separating the bones and joints. The apparently cleaner output from SVM can actually mislead the user in thinking this is accurate.

---

[1] Source code for this work can be found at https://github.com/naimulkhan/SSVMVolRen

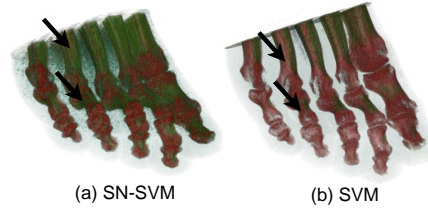

(a) SN-SVM  (b) SVM

Fig. 5. Comparison between SN-SVM and SVM.

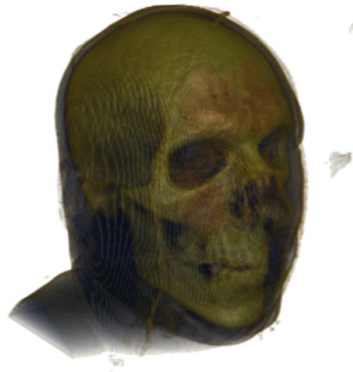

Fig. 6. Rendering results for the Visual Male Head dataset.

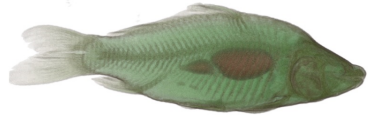

Fig. 7. Rendering results for the Carp dataset.

The rendering results for the other datasets are shown in Figure 6, 7 and 8. All these rendering results were achieved by defining two voxel classes with our GUI. We can see that like the data-centric approach [Khan et al. 2014b; Khan et al. 2015], the image-centric method can provide effective rendering results on these datasets easily and efficiently.

Table I lists the volume dimensions, number of training samples, and times required for the SN-SVM method (both training and classification). This table reveals a key strength of the image-centric approach. We can see that with the image-centric approach, rendering results can be obtained very quickly without any pre-processing. If we compare the total time with the data-centric approach [Khan et al. 2014b; Khan et al. 2015], we can see that with the data-centric approach, a dataset has to go through the training process at least once, which can take significant amount of time. But with our image-centric approach, a new dataset can be loaded immediately and worked on to achieve satisfactory results quickly and efficiently. We also see the power of the proposed SN-SVM method. From Table I it can be seen that compared to the whole volume, the number of training voxels is very small. Due to combining the local and global information from the training data, the SN-SVM method can provide satisfactory results even with such a small number of training samples.

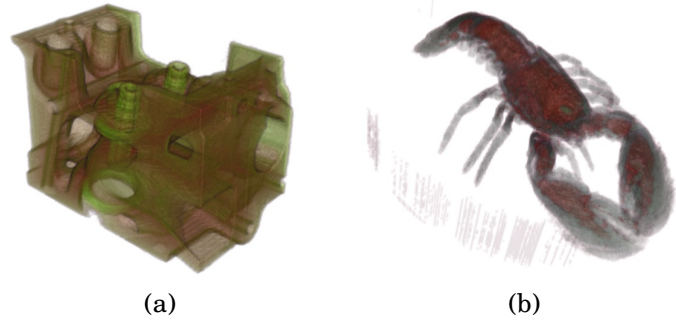

Fig. 8. (a) Rendering results for the Engine dataset. (b)Rendering results for the Lobster dataset.

Table I. Dataset details and required times (in *seconds*).

| Dataset | Dimension | # Training Samples | Total Time |
|---|---|---|---|
| Foot | 256X256X256 | 401 | 12.33 |
| Visual Male Head | 128X256X256 | 233 | 2.27 |
| Carp | 256X256X512 | 299 | 4.62 |
| Engine | 256X256X256 | 180 | 1.07 |
| Lobster | 301X324X56 | 97 | 0.67 |

**4.2. Comparative Evaluation**

The data-centric and image-centric approaches are different in principle. As discussed before, volume visualization is mostly a qualitative method. The most important aspect is the user's satisfaction. To further strengthen the usability of our system, we conducted a user survey, where we compared our proposed image-centric approach with the data-centric approach proposed by [Khan et al. 2014b; Khan et al. 2015] and an off-the-shelf traditional TF editor [Kitware[TM] 2015]. A point to note here is that to our knowledge, there is no existing method to quantitatively compare volume rendering techniques, since the final interpretation is user dependent. Some measures have been proposed recently to tackle this problem [Wu et al. 2007]. Even those measurements had to undergo extensive user studies [Wu et al. 2007] to justify their applicability. The problem with evaluation of volume rendering systems lies in the fact that there are too many different parameters from system to system to come up with a fair and viable comparison platform. Most of the proposed methods make use of their own unique interfaces to make the user interaction easier and more efficient. The TF generation process also varies significantly. As a result, we have resorted to the commonly accepted user satisfaction and efficiency as comparison tools [Giesen et al. 2007; Walimbe et al. 2003].

For our user survey, 10 users were carefully chosen to represent varying levels of knowledge and familiarity with visualization tools and techniques. Users 1-5 are the most familiar with graphics tools and techniques, with some knowledge of the DVR rendering pipeline. User 6-8 have had previous experience in using complex software tools, not necessarily graphics-related. Users 9-10 have little to no experience with similar tools. There were two phases in the survey:

— **Training**: In this phase, the users were familiarized with the interface of the systems. This phase was continued till the user was comfortable with the various options (e.g. selecting voxels, defining voxel groups in our proposed system and selecting points from histogram in the traditional TF editor).
— **Evaluation**: The users were provided with the raw visualization of the Foot dataset without any voxel grouping (Figure 9-(a)) and a reference image with three voxel groups (Figure 4-(b)). The task was to play with the different options in the system and generate a similar rendering as the reference image from the raw visualization. In our proposed system, the color and opacity values are automatically generated through color harmonization. For the traditional TF editor, the user was free to choose arbitrary color and opacity values. Since the reference image was generated from our system, to be fair to the traditional method, matching the color and opacity was not a requirement, as long as all three voxel groups were distinguishable to a satisfactory level for the user.

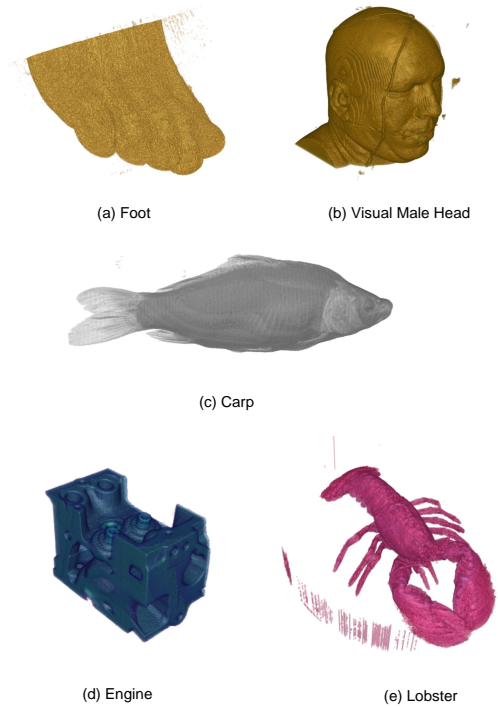

(a) Foot　　　　　　　　(b) Visual Male Head

(c) Carp

(d) Engine　　　　　　　(e) Lobster

Fig. 9. Intermediate rendering of all the datasets without any voxel classification or color harmonization.

During these two phases, the time required was recorded for each user (Table II). After the evaluation, the users were asked to rate each system on a Likert scale [Likert 1932] of 5, with 5 being the best (completely satisfied) and 1 being the worst (not satisfied). There were two categories for rating, the interaction with the system and the quality of the output.

Table II summarizes the results from this user survey. As we can see, the image-centric and data-centric approaches perform significantly better in all aspects. For the training phase, we see that all users needed more time to adapt to the traditional TF editor than to our proposed system. The time difference is especially significant in case of Users 9-10, who happens to be the least experienced with visualization systems. To interact with the traditional TF editor, you need some knowledge of histogram and color/opacity mapping. Since these users were not familiar with these concepts, it took them more time to adapt. However, our proposed system is simple and intuitive. No low level parameter is exposed to the user. As a result, the training for our system was faster. Even with other experienced users (1-8), we see that the training of our proposed system was faster.

The user survey (Table II) revealed some interesting pros and cons of both the data-centric and image-centric approaches. We can see that in general, the image-centric method is much faster in terms of both training and interaction time. This is because the image-centric method is simpler. The user directly works in the image domain, there is no need to explain any clustering or histogram techniques.

However, we can see that the users rated the image-centric approach much lower than the data-centric one. With the data-centric approach, we can cluster the voxel features and show interesting regions in the cluster domain. Due to high-dimensional clustering, this domain might be able to convey the information available in the volume more concisely. In the image domain, the user typically works on grayscale representation of the volume slices. Although in later stages higher-dimensional features are being used for classification, the training data is selected from the slice representations only. The user might miss some regions that are not apparent until a higher-dimensional clustering is done. Also, the users in general felt that the image-centric approach did not give them the granular level of control that the data-centric approach could provide. In the data-centric approach, they could immediately see which voxels were being selected before the voxel group was defined. But in the image-centric approach, the render was only shown after the voxel group definition is finalized. To summarize, the obvious advantage of image-centric approach is the speed and efficiency in generating a render. But the render might not be as good in quality as can be done in the data-centric approach.

## 5. CONCLUSION

In this paper we have proposed a novel image-centric volume visualization approach where the user directly interacts with the data to select interesting structures. Treating the user input as training data, the SN-SVM classifier combined with the concept of color harmonization can generate accurate output showing easily distinguishable structures with aesthetically pleasing colors. Experimental results on several datasets have shown the effectiveness and efficiency of the system. Compared to data-centric approaches, the obvious advantage of image-centric approach is the speed and efficiency in generating a render.

The system has a few limitations that will be worked on in future. The datasets used here are not necessarily very complex. For more complex datasets, it is likely that the features used in the proposed system will have to be more sophisticated. However, for a proof-of-concept system, the results, especially the user survey shows that the system has strong potential. Also, due to the modular design of the proposed system, adapting it to a more sophisticated feature set will require minimal to no changes to other aspects of the system i.e. SN-SVM and rendering.

We will also work on devising an efficient post-classification rendering process for the proposed system so that the rendering quality of the output volume images get better.

Table II. Times recorded for user training and interaction (in *seconds*) and the ratings (on a scale of 5) by the users for interaction and output quality.

| User | System | Training Time | Interaction Time | Interaction Rating | Output Rating |
|---|---|---|---|---|---|
| User 1 | Data-centric | 110 | 130 | 4.5 | 4 |
| | Image-centric | 30 | 15 | 3.5 | 3 |
| | Traditional | 170 | 250 | 3.5 | 2.5 |
| User 2 | Data-centric | 90 | 125 | 4 | 5 |
| | Image-centric | 25 | 17 | 3.5 | 3 |
| | Traditional | 150 | 190 | 3 | 3 |
| User 3 | Data-centric | 120 | 130 | 4 | 4 |
| | Image-centric | 39 | 21 | 3.5 | 3.5 |
| | Traditional | 177 | 200 | 3.5 | 3.5 |
| User 4 | Data-centric | 107 | 140 | 4.5 | 4 |
| | Image-centric | 20 | 20 | 4 | 4 |
| | Traditional | 150 | 220 | 4 | 3.5 |
| User 5 | Data-centric | 100 | 100 | 5 | 4 |
| | Image-centric | 35 | 13 | 4 | 3 |
| | Traditional | 142 | 198 | 3 | 3 |
| User 6 | Data-centric | 100 | 130 | 5 | 4.5 |
| | Image-centric | 40 | 21 | 4 | 3.5 |
| | Traditional | 210 | 340 | 3.5 | 3 |
| User 7 | Data-centric | 140 | 140 | 4.5 | 4.5 |
| | Image-centric | 44 | 29 | 4 | 3 |
| | Traditional | 220 | 310 | 3.5 | 2.5 |
| User 8 | Data-centric | 125 | 105 | 4 | 4 |
| | Image-centric | 30 | 31 | 3.5 | 3.5 |
| | Traditional | 190 | 350 | 4 | 3.5 |
| User 9 | Data-centric | 170 | 220 | 4 | 5 |
| | Image-centric | 50 | 39 | 4 | 3.5 |
| | Traditional | 580 | 610 | 3 | 2.5 |
| User 10 | Data-centric | 150 | 200 | 4.5 | 5 |
| | Image-centric | 47 | 41 | 4 | 3 |
| | Traditional | 450 | 570 | 2.5 | 2.5 |
| Average | Data-centric | 121.2 | 158 | 4.4 | 4.4 |
| | Image-centric | 36 | 24.7 | 3.8 | 3.3 |
| | Traditional | 243.9 | 323.8 | 3.35 | 2.95 |